\documentclass[
  reprint,
  amsmath,amssymb,
  aps,
  prb,
  floatfix,
  superscriptaddress,
]{revtex4-2}

\usepackage{graphicx}
\usepackage{dcolumn}
\usepackage{bm}
\usepackage[hidelinks]{hyperref}

\begin{document}

\title{Dynamic backaction in ferromagnet--superconductor hybrids: Coupled micromagnetic and Ginzburg--Landau modeling of ferromagnetic resonance}

\author{Qingqin Dai}
\affiliation{State Key Laboratory of Surface Physics and Institute for Nanoelectronic Devices and Quantum Computing, Fudan University, Shanghai 200433, China}
\affiliation{Department of Physics, Fudan University, Shanghai 200433, China}

\author{Weichao Yu}
\email[Corresponding author: ]{wcyu@fudan.edu.cn}
\affiliation{State Key Laboratory of Surface Physics and Institute for Nanoelectronic Devices and Quantum Computing, Fudan University, Shanghai 200433, China}
\affiliation{Zhangjiang Fudan International Innovation Center, Fudan University, Shanghai 201210, China}

\begin{abstract}
We develop a self-consistent numerical model coupling the Landau--Lifshitz--Gilbert and time-dependent Ginzburg--Landau equations through the electromagnetic field. Applying it to ferromagnetic resonance of a single-domain sphere above a superconducting plane in the Meissner state, we find that static screening alone predicts a redshift, whereas dynamic backaction produces a blueshift. Finite penetration is captured approximately by an effective image plane shifted into the superconductor. For a pinned vortex, the vortex and screening fields decay differently with distance, producing monotonic or nonmonotonic frequency--height relations depending on sphere radius. This framework can be extended to nonuniform magnetic modes, multivortex dynamics, and arbitrary ferromagnet--superconductor geometries, providing a basis for bidirectional control of magnetic excitations and superconducting states in these hybrids.
\end{abstract}
\maketitle

\section{Introduction}
\label{sec:introduction}

Ferromagnet--superconductor hybrids couple magnetic order to superconductivity through electronic proximity effects or stray-field-mediated electromagnetic interactions, providing versatile platforms to control spin transport and magnetic excitations~\cite{Lyuksyutov2005Hybrids,Buzdin2005Proximity,Linder2015Spintronics,Yu2026Electromagnetic,LoConte2024MagnetSuperconductor}. In this work, we focus on hybrids without direct electronic contact, where the coupling is mediated entirely by stray fields. The physical essence of this magnetostatic coupling is inherently bidirectional. The stray field of a ferromagnet induces screening supercurrents and modifies the vortex distribution in the superconductor~\cite{Morgan1998AsymmetricFluxPinning}. In return, the magnetic field generated by these supercurrents and vortices feeds back on the ferromagnet, altering its magnetization dynamics. This mutual interaction implies that a complete understanding of how a superconductor controls magnetism, and conversely how magnetism controls superconductivity, requires a numerical model that treats both subsystems on equal footing.

This bidirectional interplay has been explored extensively across static and dynamic regimes. Regarding the superconducting control of magnetism, screening and dynamic electromagnetic feedback have been shown to modify spin-wave dispersion, magnon transport, and thin-film ferromagnetic resonance (FMR), with theoretical treatments accounting for finite-penetration and demagnetizing-field corrections~\cite{Golovchanskiy2018Hybridization,Golovchanskiy2018Dispersion,Golovchanskiy2019Metamaterials,Mironov2021Demagnetization,Silaev2022AndersonHiggs,Zhou2023Gating,Popadiuk2026Mechanism,Yu2022Gating,Zhou2024Transport,Kharlan2024Confinement,Kharlan2024MagnonicCrystal}. Experiments have further revealed substantial shifts in FMR across the superconducting transition~\cite{Jeon2019Screening,Golovchanskiy2020Dynamics,Golovchanskiy2023Torque,Tian2025Meissner,Li2025FMRGating} and the emergence of hybrid spin-wave--Meissner-current modes~\cite{Borst2023Hybrid}. Conversely, the magnetic subsystem can profoundly affect the superconducting state. Static magnetic textures, such as domain walls and skyrmions, can template or pin superconducting vortices, leading to the formation of skyrmion-vortex pairs~\cite{Petrovic2021SkyrmionAntivortex,Neto2022SkyrmionVortexMatter,Xie2024SkyrmionVortexPairs} and complex magnon-fluxon interactions~\cite{Katkov2024BoundStates,Dobrovolskiy2019MagnonFluxon,Dobrovolskiy2025VortexMagnons}. Furthermore, magnetization dynamics can induce alternating supercurrents, drive vortex motion, and change superconducting transport~\cite{Milosevic2004VortexAntivortex,Carapella2010Asymmetry,Jing2014Vortex,Wilcox2025MagneticallyControlledVortex}. Because both static stray fields and alternating electromagnetic feedback are essential to these phenomena, a model of the bidirectional coupling must account for both. However, treating the two effects self-consistently within a single model remains a central challenge.

The methods used to study these phenomena can be grouped by how the coupling is closed. In unidirectional treatments one subsystem is prescribed. Either the superconducting state is fixed as a screening background or a prescribed vortex configuration and the magnetic response is computed~\cite{Golovchanskiy2018Hybridization,Golovchanskiy2019Metamaterials,Mironov2021Demagnetization,Niedzielski2023MagnonFluxon,Katkov2024BoundStates}, or the magnetic drive is prescribed and the superconducting response is computed~\cite{jafri_numerical_2017,Oripov2020TDGL,Milosevic2004VortexAntivortex,Carapella2010Asymmetry,Jing2014Vortex}. Jafri et al.\ addressed the coupled problem early on with an in-house three-dimensional code coupling the Landau--Lifshitz--Gilbert (LLG) equation~\cite{Gilbert2004Damping} to the time-dependent Ginzburg--Landau (TDGL) equations~\cite{Schmid1966TDGL}, which reproduced domain-templated vortex patterns and transport~\cite{jafri_domain_2020,mehdi_jafri_phase-field_2020}. In that scheme the two subsystems are advanced in sequence. The ferromagnetic domain pattern, including nonuniform textures, is first relaxed with the LLG equation and then held fixed as a static stray-field template while the TDGL equations are evolved. This is well matched to the equilibrium configurations and transport addressed there, but the magnetization does not respond to the superconducting state it creates. Conversely, the closed loop in which magnetization dynamics induces an alternating supercurrent whose field reacts back on the magnetization has been established for magnon gating~\cite{Yu2022Gating}, for Meissner-enhanced magnon transport~\cite{Zhou2024Transport}, and for the FMR shift of superconductor/ferromagnet/superconductor trilayers~\cite{Popadiuk2026Mechanism}, with related dynamic-image and effective-screening treatments~\cite{Golovchanskiy2018Dispersion,Silaev2022AndersonHiggs,Zhou2023Gating}. There the superconductor is replaced by an effective boundary or radiation response, tailored to the high-symmetry geometries for which it was derived. Each line thus freezes what the other keeps dynamical. A real-space framework in which neither subsystem is frozen, together with its frequency-domain counterpart for linear-response spectroscopy such as FMR, has to our knowledge not been established.

Here we develop a coupled micromagnetic and Ginzburg--Landau framework in which the LLG equation for the magnetization and the TDGL equations for the superconductor interact bidirectionally and self-consistently through the electromagnetic field in both the time and frequency domains. We use this model to examine how dynamic backaction affects local FMR in the presence of finite penetration and a pinned vortex. The paper is organized as follows. Section~\ref{sec:model} presents the model and numerical methods. Section~\ref{sec:results} examines the Meissner and pinned single-vortex states, including the roles of screening and vortex stray fields and the dependence on sphere radius and separation. Section~\ref{sec:discussion} discusses the scope of the model, possible experimental tests, and extensions. Section~\ref{sec:conclusion} concludes the paper.

\section{Model}
\label{sec:model}

\begin{figure}[t]
  \centering
  \includegraphics[width=\columnwidth]{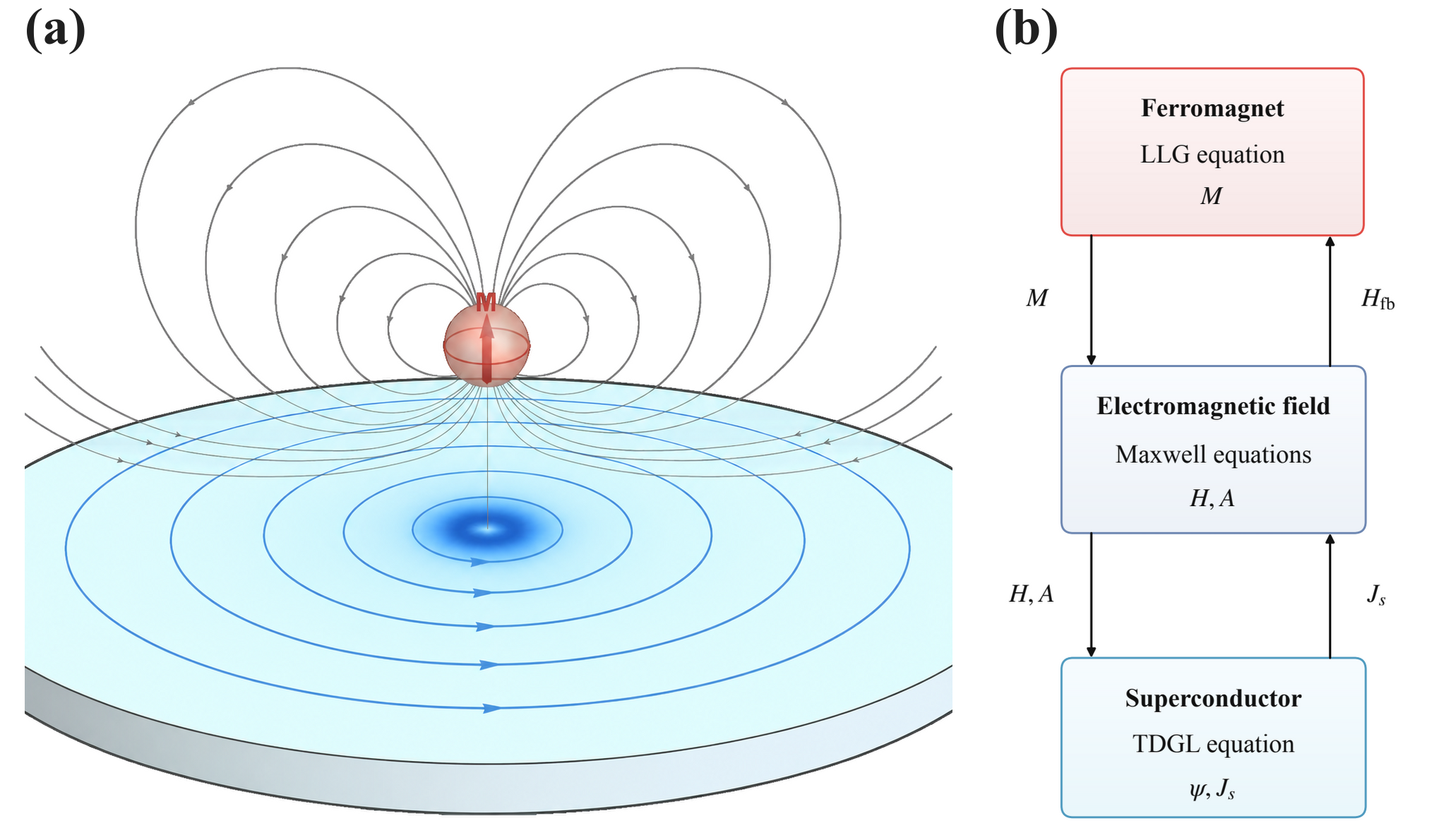}
  \caption{Bidirectional coupling between a ferromagnet and a superconductor through stray fields. (a) A ferromagnetic sphere above a superconducting plane. Its stray field induces supercurrents that generate a feedback field. The equilibrium magnetization is in-plane (along $x$) for the Meissner-state calculations and out-of-plane (along $z$) for the vortex-state calculations. (b) Coupling in the numerical model: the magnetization $\mathbf{M}$ is obtained from the LLG equation [Eq.~(\ref{eq:llg})], while the superconducting order parameter $\Psi$ and supercurrent $\mathbf{J}_s$ are described by the TDGL equations [Eqs.~(\ref{eq:tdgl_psi}) and (\ref{eq:tdgl_A})]. The two subsystems interact through the stray electromagnetic field obtained from Maxwell's equations.}
  \label{fig:model}
\end{figure}

The bidirectionally coupled ferromagnet--superconductor system considered here is shown in Fig.~\ref{fig:model}(a). A ferromagnetic sphere is placed above the planar surface of a sufficiently thick superconductor. The coupling is mediated entirely by stray fields, with no direct electronic contact between the two subsystems.

Precession of the magnetization produces a time-dependent stray field that induces supercurrents near the superconducting surface. The field generated by these currents, in turn, modifies the local field at the sphere and hence its magnetization dynamics.
The magnetization and superconducting screening response thus form the closed feedback loop referred to here as bidirectional coupling.

\subsection{LLG equation}
\label{sec:llg}

We treat the ferromagnetic sphere as a single-domain magnet. Its equilibrium magnetization profile is obtained self-consistently by relaxing the LLG equation under the combined influence of the external bias field and the superconducting stray field. The state of the sphere is described by the normalized magnetization $\mathbf{m}=\mathbf{M}/M_s$, where $\mathbf{M}$ is the magnetization and $M_s$ is the saturation magnetization.

The magnetization obeys the Landau--Lifshitz--Gilbert equation~\cite{Gilbert2004Damping},
\begin{equation}
\frac{\partial \mathbf{m}}{\partial t}
=
-\gamma \mathbf{m}\times \mathbf{H}_{\mathrm{eff}}
+\alpha_{\mathrm G} \mathbf{m}\times
\frac{\partial \mathbf{m}}{\partial t},
\label{eq:llg}
\end{equation}
where $\gamma>0$ is the magnitude of the gyromagnetic ratio, $\alpha_{\mathrm G}>0$ is the Gilbert damping constant, and $\mathbf{H}_{\mathrm{eff}}$ is the effective magnetic field.
The first term describes precession about the effective field. The second accounts for damping-driven relaxation toward it.

The effective field contains the applied static field $\mathbf{H}_0$ and the dipolar (stray) field $\mathbf{H}_{\mathrm{D}}$, which comprises the sphere's own demagnetizing field and the feedback field returned by the superconductor:
\begin{equation}
\mathbf{H}_{\mathrm{eff}}
=
\mathbf{H}_0+\mathbf{H}_{\mathrm{D}}.
\label{eq:heff}
\end{equation}

Within the ferromagnet, the magnetic flux density satisfies
\begin{equation}
\mathbf{B}=\mu_0(\mathbf{H}_{\mathrm{D}}+\mathbf{M}).
\label{eq:b_fm}
\end{equation}
The magnetization $\mathbf{M}$ therefore enters the spatial field calculation as a magnetic source, and the resulting stray field $\mathbf{H}_{\mathrm{D}}$ is fed back into the LLG equation [Eq.~(\ref{eq:llg})]. To leading order, the self-demagnetizing part of $\mathbf{H}_{\mathrm{D}}$ is $-\mathbf{M}/3$, parallel to $\mathbf{m}$, and exerts no net torque. The precession is governed by $\mathbf{H}_0$ and the superconducting contribution to $\mathbf{H}_{\mathrm{D}}$. The bias $\mathbf{H}_0$ is not a source of the field in Eq.~(\ref{eq:b_fm}), which is produced only by the magnetization $\mathbf{M}$ and by the currents flowing in the superconductor. It enters separately as a uniform term in the effective field of Eq.~(\ref{eq:heff}) to stabilize the magnetization along the chosen direction and set the precession frequency. The effect of the static bias field on the superconductor is assumed to be weak and is therefore neglected in the TDGL modeling.

\subsection{TDGL equations}
\label{sec:tdgl}

We describe the superconducting response with the TDGL equations~\cite{Schmid1966TDGL,Cyrot1973GLTheory}, which extend the static Ginzburg--Landau formalism to nonequilibrium evolution. These equations have been solved numerically with finite-element and finite-difference schemes for vortex entry in samples of complex geometry and for vortex dynamics in oscillating magnetic fields~\cite{Alstrom2011TDGLFEM,jafri_numerical_2017}. Three-dimensional TDGL simulations have previously treated superconductors driven by a local magnetic dipole~\cite{Oripov2020TDGL}. Here the source is itself dynamic. The precessing magnetization induces an alternating supercurrent $\delta\mathbf{J}_s$, whose field $\delta\mathbf{H}_{\mathrm{eff}}$ enters the LLG equation [Eq.~(\ref{eq:llg})] and corrects the precession frequency, closing the feedback loop.
The full TDGL model comprises coupled evolution equations for the superconducting order parameter $\Psi$ and magnetic vector potential $\mathbf{A}$.

In the zero-scalar-potential gauge, $\Phi=0$, the order parameter obeys
\begin{equation}
\frac{\hbar^2}{2m_\ast D}
\frac{\partial \Psi}{\partial t}
=
-\frac{1}{2m_\ast}
\left(
\frac{\hbar}{i}\nabla-e_\ast\mathbf{A}
\right)^2\Psi
-\alpha\Psi
-\beta|\Psi|^2\Psi.
\label{eq:tdgl_psi}
\end{equation}
Here $m_\ast=2m_e$ and $e_\ast=2|e|$ are the effective Cooper-pair mass and the magnitude of its charge, respectively, $D$ is a relaxation coefficient, and $\mathbf{A}$ is the magnetic vector potential. The Ginzburg--Landau coefficients $\alpha$ and $\beta$ determine the coherence length $\xi$ and magnetic penetration depth $\lambda$:
\begin{equation}
\xi=
\sqrt{\frac{\hbar^2}{2m_\ast|\alpha|}},
\qquad
\lambda=
\sqrt{\frac{m_\ast\beta}{\mu_0e_\ast^2|\alpha|}}.
\label{eq:xi_lambda}
\end{equation}
Their ratio defines the Ginzburg--Landau parameter,
\begin{equation}
\kappa=\frac{\lambda}{\xi},
\label{eq:kappa}
\end{equation}
which distinguishes type-I superconductivity, $\kappa<1/\sqrt{2}$, from type-II superconductivity, $\kappa>1/\sqrt{2}$.

The equation for the vector potential is
\begin{align}
\sigma
\frac{\partial \mathbf{A}}{\partial t}
={}&
\frac{e_\ast\hbar}{2m_\ast i}
\left(
\Psi^\ast\nabla\Psi-\Psi\nabla\Psi^\ast
\right)
\nonumber\\
&-\frac{e_\ast^2}{m_\ast}|\Psi|^2\mathbf{A}
-\frac{1}{\mu_0}\!
\nabla\times
\left(
\nabla\times\mathbf{A}-\mathbf{B}_a
\right).
\label{eq:tdgl_A}
\end{align}
Here $\sigma$ is the normal-state electrical conductivity, and \(\mathbf{B}_a = \mu_0\mathbf{M}\) acts as the source flux density. The magnetization supplied by the LLG equation enters as a source, and the resulting field is fed back into the LLG equation. 

Outside the superconductor, including inside the ferromagnetic sphere, the order parameter $\Psi$ and the conductivity $\sigma$ vanish. Consequently, Eq.~(\ref{eq:tdgl_A}) naturally reduces to the magnetostatic Amp\`ere law, \(\nabla\times\mathbf{B}=\mu_0\mathbf{J}_{\mathrm{ext}}\) with \(\mathbf{B}=\nabla\times\mathbf{A}\) and \(\mathbf{J}_{\mathrm{ext}}=\nabla\times\mathbf{M}\). 

Therefore, Eq.~(\ref{eq:tdgl_A}) provides a unified description across the entire simulation domain. The vector potential $\mathbf{A}$ continuously connects the superconducting and non-superconducting regions, and this global equation closes the self-consistent loop between the LLG and TDGL subsystems.

Comparison with the generalized Amp\`ere law identifies the supercurrent as
\begin{equation}
\mathbf{J}_s
=
\frac{e_\ast\hbar}{2m_\ast i}
\left(
\Psi^\ast\nabla\Psi-\Psi\nabla\Psi^\ast
\right)
-\frac{e_\ast^2}{m_\ast}
|\Psi|^2\mathbf{A}.
\label{eq:supercurrent}
\end{equation}
The supercurrent density $\mathbf{J}_s$ is thus determined jointly by the order parameter and vector potential. This feedback closes the self-consistent loop illustrated in Fig.~\ref{fig:model}(b).

\subsection{Frequency-domain formulation}
\label{sec:frequency_domain}

The preceding equations are formulated in the time domain and can treat arbitrary dynamical problems. For linear-response behavior such as FMR, a frequency-domain calculation is faster and more convenient. The frequency-domain form of the linearized LLG equation was established by Zhang et al.~\cite{Zhang2023FrequencyDomain} and subsequently applied in frequency-domain micromagnetic simulations of magnon-polaron hybridization~\cite{Wang2026MagnonPolaron}. We extend this frequency-domain framework by adding the TDGL equations and their coupling to the LLG equation, introducing small harmonic perturbations about a static equilibrium and linearizing the coupled equations.

For the ferromagnet, we follow this frequency-domain treatment of the linearized LLG equation about equilibrium and write
\begin{equation}
\begin{aligned}
\mathbf m&=\mathbf m_0+\delta\mathbf m e^{i\omega t},\\
\mathbf H_{\mathrm{eff}}&=\mathbf H_{\mathrm{eff},0}
+\delta\mathbf H_{\mathrm{eff}}e^{i\omega t}.
\end{aligned}
\label{eq:frequency_llg_perturbation}
\end{equation}
The subscript $0$ denotes the equilibrium state and $\delta$ a first-order perturbation. Substituting Eq.~(\ref{eq:frequency_llg_perturbation}) into Eq.~(\ref{eq:llg}) and retaining linear terms gives
\begin{equation}
\begin{aligned}
i\omega\delta\mathbf m
={}&-\gamma\left(
\mathbf m_0\times\delta\mathbf H_{\mathrm{eff}}
+\delta\mathbf m\times\mathbf H_{\mathrm{eff},0}
\right)\\
&+i\alpha_{\mathrm G}\omega\,
\mathbf m_0\times\delta\mathbf m.
\end{aligned}
\label{eq:frequency_llg}
\end{equation}

For the superconductor, we similarly write
\begin{equation}
\begin{aligned}
\Psi&=\Psi_0+\delta\Psi e^{i\omega t},\\
\mathbf A&=\mathbf A_0+\delta\mathbf A e^{i\omega t}.
\end{aligned}
\label{eq:harmonic_perturbation}
\end{equation}
Substitution of Eq.~(\ref{eq:harmonic_perturbation}) into Eq.~(\ref{eq:tdgl_psi}) gives the linearized frequency-domain equation,
\begin{widetext}
\begin{equation}
\begin{aligned}
i\omega\frac{\hbar^2}{2m_\ast D}\delta\Psi
={}&\frac{\hbar^2}{2m_\ast}\nabla^2\delta\Psi
-\alpha\delta\Psi
-\beta\bigl(2|\Psi_0|^2\delta\Psi+\Psi_0^2\delta\Psi^\ast\bigr)\\
&-\frac{i\hbar e_\ast}{2m_\ast}\bigl[(\nabla\cdot\mathbf A_0)\delta\Psi
+2\mathbf A_0\!\cdot\!\nabla\delta\Psi
+(\nabla\cdot\delta\mathbf A)\Psi_0
+2\delta\mathbf A\!\cdot\!\nabla\Psi_0\bigr]
-\frac{e_\ast^2}{2m_\ast}\bigl[|\mathbf A_0|^2\delta\Psi
+2(\mathbf A_0\!\cdot\!\delta\mathbf A)\Psi_0\bigr].
\end{aligned}
\label{eq:frequency_tdgl}
\end{equation}
Here $\delta\Psi^\ast\equiv[\delta\Psi(-\omega)]^\ast$ denotes the phasor of the conjugate order-parameter perturbation at frequency $+\omega$, i.e., the complex conjugate of the $-\omega$ component. It is not the complex conjugate of the $+\omega$ phasor $\delta\Psi$. Equivalently, the real and imaginary parts of $\Psi$ are linearized as independent real fields.

Expanding the supercurrent in Eq.~(\ref{eq:supercurrent}) to first order gives
\begin{equation}
\delta\mathbf J_s
=\frac{e_\ast\hbar}{2m_\ast i}\bigl[\Psi_0^\ast\nabla\delta\Psi
+\delta\Psi^\ast\nabla\Psi_0
-\Psi_0\nabla\delta\Psi^\ast
-\delta\Psi\nabla\Psi_0^\ast\bigr]
-\frac{e_\ast^2}{m_\ast}\bigl[|\Psi_0|^2\delta\mathbf A
+(\Psi_0^\ast\delta\Psi+\Psi_0\delta\Psi^\ast)\mathbf A_0\bigr].
\label{eq:frequency_supercurrent_full}
\end{equation}
\end{widetext}
The alternating supercurrent contains contributions from both the order-parameter perturbation $\delta\Psi$ and vector-potential perturbation $\delta\mathbf A$. Equation~(\ref{eq:frequency_supercurrent_full}), solved together with Eq.~(\ref{eq:frequency_tdgl}) for $\delta\Psi$, could in principle be used directly to close the bidirectional loop. In practice, we neglect the order-parameter perturbation in the current response,
\begin{equation}
\delta\Psi\simeq0,\qquad\delta\Psi^\ast\simeq0.
\label{eq:frequency_fixed_psi}
\end{equation}
This neglect is justified by the structure of the source terms in Eq.~(\ref{eq:frequency_tdgl}). In the zero-scalar-potential gauge, charge neutrality makes the alternating electric field transverse, $\nabla\cdot\delta\mathbf{A}=0$, so $\delta\Psi$ is driven only by terms proportional to $\nabla\Psi_0$ and $\mathbf{A}_0$. In the Meissner bulk $\Psi_0$ is uniform and $\mathbf{A}_0$ decays within the penetration depth, so $\delta\Psi$ is confined to the surface layer and, in the vortex state, to the core. Resolving it would require a dedicated mesh in these thin regions. Freezing $\delta\Psi$ thus improves numerical stability at negligible cost to the bulk current response. The current perturbation then reduces to
\begin{equation}
\delta\mathbf J_s
=-\frac{e_\ast^2}{m_\ast}|\Psi_0|^2\delta\mathbf A.
\label{eq:frequency_supercurrent}
\end{equation}
Here $\Psi_0$ is the dimensional equilibrium order parameter. Equation~(\ref{eq:frequency_fixed_psi}) freezes only the dynamical component of the order parameter. The alternating supercurrent of Eq.~(\ref{eq:frequency_supercurrent}), and hence the superconducting feedback field entering the LLG equation, survives this freezing, so the bidirectional coupling remains intact.

Earlier work obtained a vortex background from the TDGL equations and then calculated spin waves in that background~\cite{Niedzielski2023MagnonFluxon}. Here we additionally solve self-consistently for the alternating currents and their feedback field on a fixed superconducting equilibrium background. The unidirectional calculation includes only the static screening background. The bidirectional calculation also captures the alternating superconducting electromagnetic response, allowing its effect on FMR to be isolated.

\subsection{Numerical simulations and parameter settings}
\label{sec:parameters}

The coupled model is implemented in COMSOL Multiphysics~\cite{comsol}. The LLG equation is solved with the Micromagnetics Module~\cite{Zhang2023FrequencyDomain,Wang2026MagnonPolaron}, and a mathematics interface solves for the superconducting order parameter. Rather than solving the full vector-potential evolution equation separately in that interface, we use the Magnetic Fields (mf) interface to calculate the spatial magnetic field, with the magnetization $\mathbf{M}$ as a magnetic source and the supercurrent density $\mathbf{J}_s$ from Sec.~\ref{sec:tdgl} as an external current density. The resulting stray field $\mathbf{H}_{\mathrm{D}}$, together with the applied field $\mathbf{H}_0$, forms the effective field of the LLG equation [Eq.~(\ref{eq:heff})] and closes the self-consistent feedback loop.

For the frequency-domain calculation, we hold the equilibrium order parameter fixed as in Sec.~\ref{sec:frequency_domain}. The reduced alternating current $\delta\mathbf{J}_s$ is supplied to the mf interface to calculate the feedback field, which is coupled to the frequency-domain LLG equation [Eq.~(\ref{eq:frequency_llg})]. When the normalized equilibrium order parameter $\psi_0$ is used in the simulation, it is converted to the dimensional equilibrium order parameter $\Psi_0$ by
\begin{equation}
\psi_0=\frac{\Psi_0}{\sqrt{-\alpha/\beta}},
\qquad
|\Psi_0|^2=-\frac{\alpha}{\beta}|\psi_0|^2,
\label{eq:psi_normalization}
\end{equation}
where $\sqrt{-\alpha/\beta}$ is the equilibrium amplitude of the order parameter in the uniform superconducting state of Eq.~(\ref{eq:tdgl_psi}). The second relation supplies the dimensional density $|\Psi_0|^2$ that enters Eq.~(\ref{eq:frequency_supercurrent}).

The numerical geometry consists of a ferromagnetic sphere above the central axis of a superconducting disk [Fig.~\ref{fig:model}(a)], with the disk radius much larger than the sphere radius so as to approximate a planar superconductor. For the Meissner-state calculations the sphere radius is $R=0.5~\mu\mathrm{m}$, and for the vortex-state calculations $R=0.1$, $0.2$, or $0.3~\mu\mathrm{m}$. In both cases the sphere-center height above the surface varies over $h=0.5$--$2.0~\mu\mathrm{m}$. The disk radius and thickness differ among the ideal-diamagnet, type-I, and type-II cases and are specified in Secs.~\ref{sec:meissner} and~\ref{sec:vortex}.

Unless otherwise stated, the simulations use the following material parameters. For the ferromagnet, we take yttrium iron garnet (YIG) with $M_s=1.94\times10^5~\mathrm{A/m}$~\cite{Anderson1964YIG}, $\gamma=2.21\times10^5~\mathrm{m/(A\,s)}$, and $\alpha_{\mathrm G}=5\times10^{-3}$. The bias field is $H_0=3.88\times10^4~\mathrm{A/m}$, applied along $x$ (in-plane) for the Meissner-state calculations and along $z$ (out-of-plane) for the vortex-state calculations. The superconducting parameters are specified by the penetration depth $\lambda$ and coherence length $\xi$. For the type-I case, we choose representative values $\lambda=20~\mathrm{nm}$ and $\xi=50~\mathrm{nm}$, giving $\kappa=0.4$. These lengths are of the same order as those of type-I materials such as Pb~\cite{Gasparovic1970Lead}. For the type-II case, $\lambda=80~\mathrm{nm}$ and $\xi=20~\mathrm{nm}$ give $\kappa=4$, consistent with dirty-Nb parameters used in previous work~\cite{Pathirana2020CurvilinearVortex}. We set $D=10^{-2}~\mathrm{m^2/s}$ as a phenomenological TDGL relaxation parameter. A diffusion coefficient of the same order has been used in transport estimates for aluminum nanowires~\cite{Bagrets2014Noise}. The normal-state conductivity in Eq.~(\ref{eq:tdgl_A}) is taken as $\sigma=6\times10^7~\mathrm{S/m}$, a representative good-conductor value.

\section{Results}
\label{sec:results}

\subsection{Meissner state}
\label{sec:meissner}

We first consider the vortex-free Meissner state and use an ideal diamagnetic plane as an analytical benchmark. In the simulation this ideal diamagnet is represented by a disk of radius $50~\mathrm{mm}$ and thickness $10~\mathrm{mm}$, which approximates an infinite half-space. With the parameters above, the free-space resonance frequency of the isolated ferromagnetic sphere is $f_0=\gamma H_0/(2\pi)\simeq1.365~\mathrm{GHz}$~\cite{Kittel1948FMR}.
At each sphere--superconductor separation, the frequency-domain resonance is identified from the peak transverse magnetization response in a frequency sweep. Independent time-domain calculations yield resonance peaks by a fast Fourier transform (FFT), providing a consistency check of the frequency-domain solution.

\begin{figure}[t]
  \centering
  \includegraphics[width=\columnwidth]{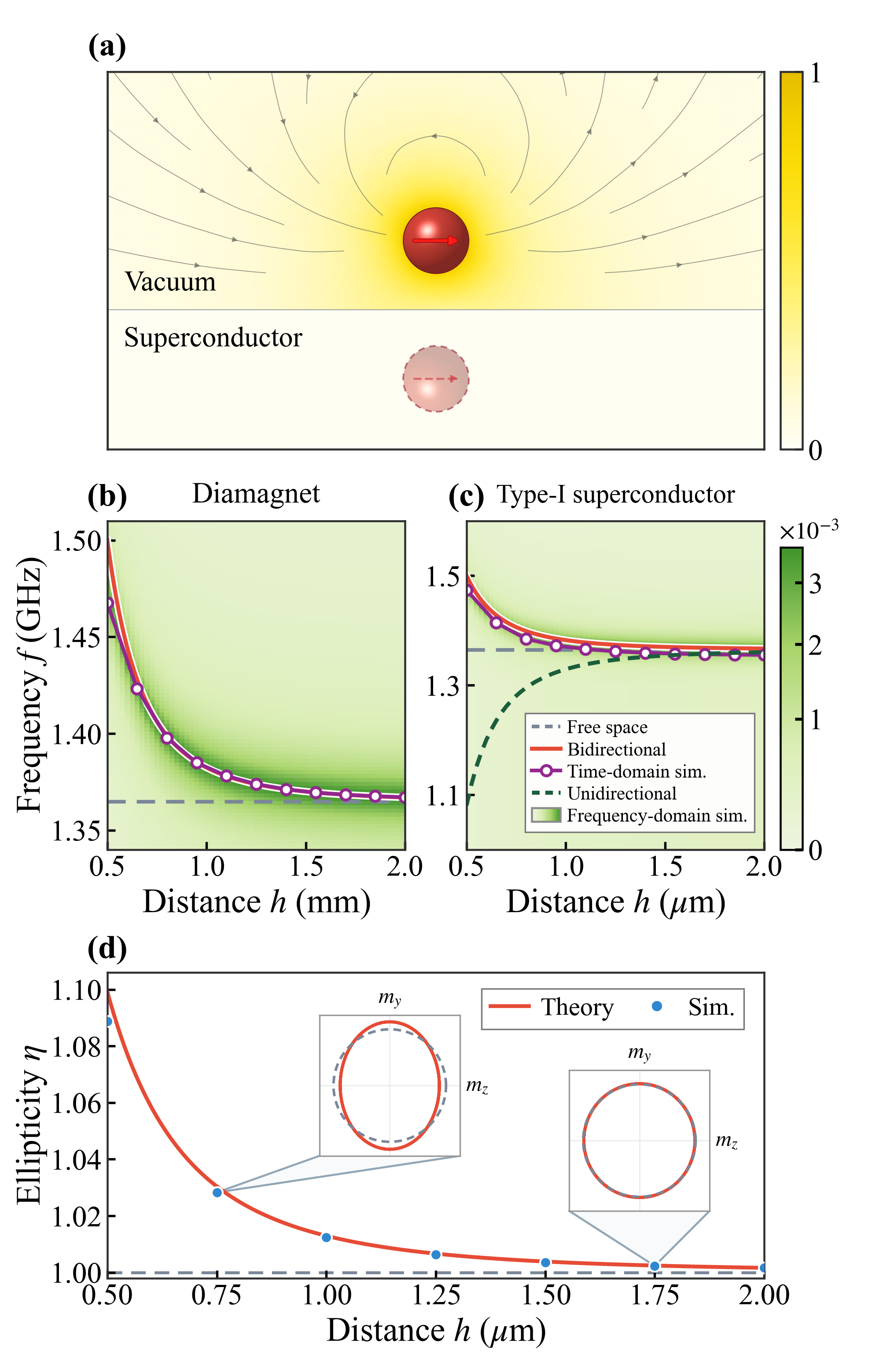}
  \caption{Concept of the image method and bidirectionally coupled numerical results in the Meissner state with in-plane magnetization. (a) Image-dipole construction for a perfectly diamagnetic plane. (b) Frequency response of a sphere above the ideal diamagnet. The color map shows the amplitude of the frequency-domain magnetization response, solid lines are image-method predictions, circles are time-domain FFT results, and the dashed line marks the free-space resonance. (c) Corresponding results for a type-I superconductor. The unidirectional result, which includes only static screening, is shown for comparison. (d) Precession ellipticity $\eta$, defined as the ratio of the $m_y$ and $m_z$ oscillation amplitudes, versus sphere-center height $h$. Lines and symbols denote the image-method and time-domain results, respectively. In both insets, deviations of the orbit from a circle are exaggerated for visibility. The main-panel values of $\eta$ are calculated directly from the original magnetization amplitudes.}
  \label{fig:meissner_benchmark}
\end{figure}

Screening by an infinite, perfectly diamagnetic plane can be described by the image method~\cite{Saslow1991SuperconductorMagnet,Kogan2003Meissner}. The physical content of this construction is that the field of the supercurrents induced in the plane is, above the surface, identical to the field of a magnetic dipole at the mirror-image position. For a sphere-center height $h$, this image dipole lies at the mirror-symmetric position below the surface, and its stray field at the sphere is
\begin{equation}
\mathbf{H}_{\mathrm{image}}
=
-H_s m_x\hat{\mathbf{x}}
-H_s m_y\hat{\mathbf{y}}
-2H_s m_z\hat{\mathbf{z}},
\label{eq:image_field_results}
\end{equation}
where
\begin{equation}
H_s(h)=\frac{\mu}{32\pi h^3}
=\frac{M_sR^3}{24h^3}.
\label{eq:Hs_results}
\end{equation}
Here $R$ is the sphere radius and $\mu=M_s(4\pi R^3/3)$ its total magnetic moment. For the Meissner-state calculations, the bias field and the equilibrium magnetization are applied in-plane (along the $x$-axis), so the transverse restoring fields for precession are governed by the $x$ and $y$ components of the image field. Substituting Eq.~(\ref{eq:image_field_results}) into the LLG equation [Eq.~(\ref{eq:llg})] and linearizing about this in-plane equilibrium gives the resonance frequency of the bidirectional calculation,
\begin{equation}
f_{\mathrm{bi}}(h)
=
\frac{\gamma}{2\pi}
\sqrt{H_0\left[H_0+H_s(h)\right]}.
\label{eq:fbi}
\end{equation}
The image field does more than shift the static bias: it acts with different coefficients on the two transverse magnetization components and therefore changes the restoring fields for precession.

To separate static screening from dynamic feedback, we also perform a unidirectional calculation where the superconductor is held in its static screening state and its alternating response is disabled. The sphere therefore experiences the static screening field, but its alternating magnetization no longer drives the superconductor. The approximate resonance frequency is then
\begin{equation}
f_{\mathrm{uni}}(h)
=
\frac{\gamma}{2\pi}\left[H_0-H_s(h)\right].
\label{eq:funi}
\end{equation}

Figure~\ref{fig:meissner_benchmark} benchmarks the numerical results against the image method. We use $R=0.5~\mu\mathrm{m}$ and sweep $h$ from $0.5$ to $2.0~\mu\mathrm{m}$. For the type-I superconductor, $\lambda=20~\mathrm{nm}$ and $\xi=50~\mathrm{nm}$ give $\kappa=0.4<1/\sqrt{2}$, placing the system in a stable Meissner state. The superconductor is modeled as a disk of radius $30~\mu\mathrm{m}$ and thickness $2~\mu\mathrm{m}$. Figure~\ref{fig:meissner_benchmark}(a) shows the ideal image configuration and field lines. Figure~\ref{fig:meissner_benchmark}(b) presents the frequency response for the ideal diamagnet, together with the image-method prediction of Eq.~(\ref{eq:fbi}) and the time-domain FFT results. The three results agree over the full separation range and approach the free-space limit $f_0$.

Figure~\ref{fig:meissner_benchmark}(c) shows the corresponding type-I results. With the full bidirectional calculation, the resonance lies above $f_0$ at short separations and agrees with both the image method and the time-domain FFT. The unidirectional result instead lies below $f_0$. The opposite shifts reflect the different physics included in the two calculations. Static diamagnetic screening lowers the effective field along the equilibrium magnetization. The bidirectional calculation additionally captures the alternating screening response driven by the alternating magnetization, which changes the transverse restoring fields and hence the precession frequency. Both screening effects weaken rapidly as $h$ increases, so the two resonance frequencies converge to the same free-space value. The bidirectional results for the ideal diamagnet and type-I superconductor nearly coincide. For the micrometer-scale geometry and parameters considered, the penetration depth is small compared with $h$, and the type-I response is close to the perfectly diamagnetic limit. Only the bidirectional calculation therefore gives the correct resonance frequency, because the dynamic backaction it retains reverses the sign of the shift rather than merely correcting it.

The anisotropy of the image field also changes the ellipticity of the magnetization precession orbit. We define the transverse ellipticity as $\eta$, which is the ratio of the oscillation amplitudes of the transverse magnetization components $m_y$ and $m_z$. Linearizing the LLG equation [Eq.~(\ref{eq:llg})] gives
\begin{equation}
\eta(h)
=
\sqrt{\frac{H_0+H_s(h)}{H_0}}.
\label{eq:eta_theory}
\end{equation}
Figure~\ref{fig:meissner_benchmark}(d) compares Eq.~(\ref{eq:eta_theory}) with time-domain simulations. At each separation, the system is evolved for $3~\mathrm{ns}$. For each complete oscillation cycle, the amplitude of each transverse component is calculated as half the difference between its maximum and minimum values. The ellipticity $\eta$ is then evaluated from these cycle-averaged amplitudes. The numerical results agree with Eq.~(\ref{eq:eta_theory}): $\eta>1$ near the surface, where precession is elliptical, and approaches 1 at larger separations, where the orbit becomes nearly circular. The insets show $m_y$--$m_z$ trajectories at two representative heights. Their deviations from a circle are exaggerated for visibility. The main-panel values of $\eta$ are calculated from the original amplitudes.

\begin{figure}[t]
  \centering
  \includegraphics[width=\columnwidth]{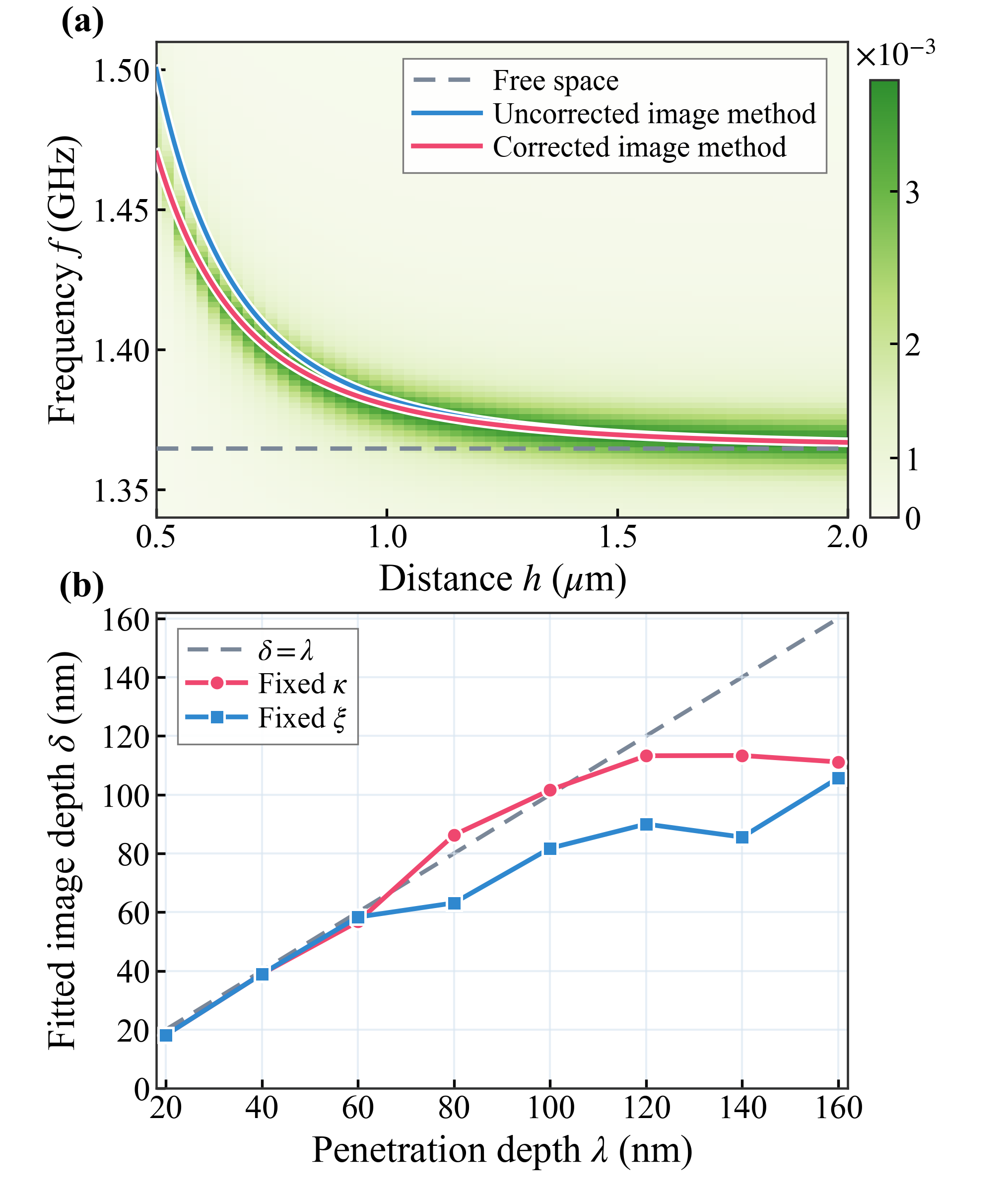}
  \caption{Finite-penetration correction to the image method. (a) Numerical frequency response and the uncorrected and corrected image-method predictions. Shifting the effective image plane into the superconductor by $\delta$ improves the description at short separations. (b) Fitted image depth $\delta$ versus penetration depth $\lambda$. Red circles correspond to fixed $\kappa=0.4$, blue squares to fixed $\xi=50~\mathrm{nm}$, and the gray dashed line to $\delta=\lambda$.}
  \label{fig:image_correction}
\end{figure}

The ideal image model confines supercurrents to the geometric surface. A finite penetration depth $\lambda$ causes the superconducting field response to depart from this limit~\cite{Kogan2003Meissner}, producing a resolvable frequency correction when $h$ reaches the micrometer scale. We denote the corrected screening field by $H_s'$. Introducing an effective image depth $\delta$, which shifts the image plane into the superconductor by $\delta$, gives
\begin{equation}
H_s'(h,R,\delta)
=
\frac{M_sR^3}{24(h+\delta)^3},
\label{eq:Hs_corrected}
\end{equation}
and the corresponding resonance frequency is
\begin{equation}
f(h,\delta)
=
\frac{\gamma}{2\pi}
\sqrt{H_0\left[H_0+H_s'(h,R,\delta)\right]}.
\label{eq:f_corrected}
\end{equation}

In Fig.~\ref{fig:image_correction}, $R=0.5~\mu\mathrm{m}$ and $h$ varies from $0.5$ to $2.0~\mu\mathrm{m}$. Figure~\ref{fig:image_correction}(a) shows a representative finite-penetration case. The uncorrected image-method prediction ($\delta=0$) systematically overestimates the shift at short separations. Allowing $\delta$ as the sole positional correction brings the theoretical curve into agreement with the numerical resonance over the full separation range. In this parameter regime, the main effect of finite penetration can therefore be represented by an inward displacement of the image plane.

To examine how $\delta$ depends on the superconducting length scales, Fig.~\ref{fig:image_correction}(b) compares two parameter paths as $\lambda$ is varied from $20$ to $160~\mathrm{nm}$. One keeps $\kappa=0.4$ fixed, so that $\xi=\lambda/\kappa$ varies with $\lambda$. The other keeps $\xi=50~\mathrm{nm}$ fixed and varies only $\lambda$. The latter deliberately crosses the type-I/type-II boundary at $\kappa=1/\sqrt{2}$. Both types support screening in the vortex-free Meissner state. This comparison therefore tests the effective image description of vortex-free screening at different Ginzburg--Landau parameters, rather than assigning all parameters to a type-I superconductor. For each parameter set, $\delta$ is fitted jointly to resonance frequencies at several values of $h$. Both paths lie close to $\delta=\lambda$ at small $\lambda$. At fixed $\kappa$, $\delta$ approximately follows $\lambda$ for $\lambda\lesssim100~\mathrm{nm}$, then departs from it and saturates near $110$--$115~\mathrm{nm}$. At fixed $\xi$, the departure occurs earlier. The effective image depth thus depends not only on $\lambda$ but also on the superconducting response controlled by $\lambda/\xi$. Accordingly, $\delta\simeq\lambda$ is an effective approximation within a finite parameter range, rather than a universal relation.

\subsection{Vortex state}
\label{sec:vortex}

We next consider a type-II superconductor containing a single quantized vortex. The superconductor is modeled as a disk of radius $6~\mu\mathrm{m}$ and thickness $1~\mu\mathrm{m}$. To hold the vortex at a prescribed position, we introduce a local pinning potential through a spatial variation of the Ginzburg--Landau coefficient:
\begin{equation}
\alpha(r)
=
\alpha_0
+
V_{\mathrm{pin}}e^{-r^2/(2R_{\mathrm{pin}}^2)}.
\label{eq:vortex_pinning}
\end{equation}
Here $r=\sqrt{(x-x_0)^2+(y-y_0)^2}$, $V_{\mathrm{pin}}=-0.5\alpha_0$, $R_{\mathrm{pin}}=20~\mathrm{nm}$, and $\alpha_0<0$. The local modification reduces the condensation-energy advantage near the pinning center and favors localization of the vortex core. The vortex-free reference uses the same parameter profile but no vortex initial condition. For the vortex calculation, a single quantum of phase winding is imposed initially, and time-domain TDGL evolution prepares the static background. The sphere is then placed above the vortex axis and its frequency response is calculated. Because the alternating current is evaluated with the equilibrium order parameter fixed, the calculation includes the static stray field of the pinned vortex and dynamic electromagnetic screening feedback, but not driven vortex displacement. 

To describe the stray field of the vortex outside the superconductor, we employ the magnetic monopole approximation~\cite{Carneiro2000VortexLines,Auslaender2009Vortices}. Unlike the closed stray field of a ferromagnetic sphere, which is dipolar, the vortex constitutes an open magnetic flux tube. Because the current-free region outside the superconductor satisfies the Laplace equation, the vortex exit point acts as a point source in the far field ($r\gg\lambda$), naturally leading to a monopolar profile~\cite{Thiel2016VortexImaging}. On the symmetry axis, replacing the radial distance $r$ by $h+h_m$ accounts for the finite spread of the screening currents over a depth of order $\lambda$. The axial field at the sphere center is therefore
\begin{equation}
H_v(h)
=
\frac{\Phi_0}
{2\pi\mu_0(h+h_m)^2},
\label{eq:Hv}
\end{equation}
where $\Phi_0$ is the flux quantum. The effective monopole depth $h_m$ captures the leading correction from the finite penetration. For the type-II parameters used here ($\lambda=80~\mathrm{nm}$, $\xi=20~\mathrm{nm}$), fitting the numerically calculated vortex field yields $h_m\simeq0.70\lambda=56~\mathrm{nm}$.

\begin{figure}[t]
  \centering
  \includegraphics[width=\columnwidth]{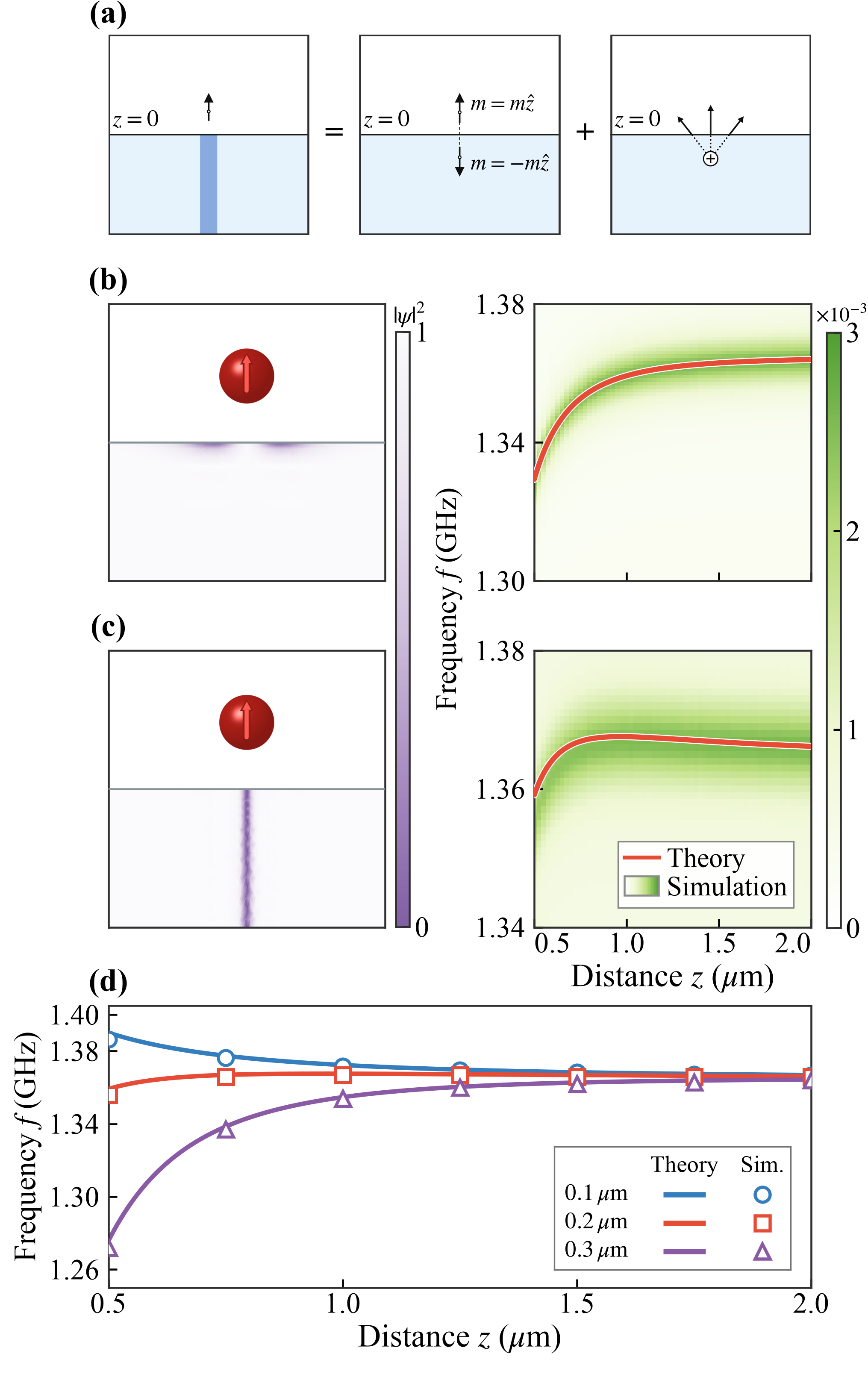}
  \caption{Competition between Meissner screening and the stray field of a single vortex under out-of-plane magnetization. (a) Analytical decomposition into an image dipole describing Meissner screening and an effective monopole describing the vortex field. (b) Vortex-free reference for $R=0.2~\mu\mathrm{m}$: $|\Psi|^2$ distribution (left) and resonance frequency versus sphere-center height with the corrected image method (middle). (c) Same as (b) after a single pinned vortex is introduced with the same magnetic parameters. The middle panel serves both (b) and (c), and its image-method curve includes both the vortex stray field and the corrected image term. (d) Resonance frequency versus sphere-center height for $R=0.1$, $0.2$, and $0.3~\mu\mathrm{m}$. Lines show Eq.~(\ref{eq:f_vortex}). Symbols are numerical results.}
  \label{fig:vortex}
\end{figure}

In Fig.~\ref{fig:vortex}, the sphere-center height $h$ varies over $0.5$--$2.0~\mu\mathrm{m}$. For the vortex-state calculations, the bias field and the equilibrium magnetization are applied out-of-plane (along the $z$-axis). The sphere experiences two superconducting contributions through stray fields: the vortex stray field $H_v$ increases the axial bias, whereas the magnet-induced Meissner screening produces an opposing corrected image field $H_s'$. Linearizing the LLG equation [Eq.~(\ref{eq:llg})] about this out-of-plane equilibrium shows that the static image field reduces the axial bias by $2H_s'$, while the transverse dynamic image response that governs the precession acts with a coefficient of 1. The resonance frequency is therefore
\begin{equation}
f(h,R)
=
\frac{\gamma}{2\pi}
\left[
H_0+H_v(h)-H_s'(h,R,\delta)
\right].
\label{eq:f_vortex}
\end{equation}
The single image coefficient in Eq.~(\ref{eq:f_vortex}) reflects the partial cancellation between the static longitudinal shift and the transverse dynamic restoring force. Equation~(\ref{eq:f_vortex}) contains two competing terms with different spatial dependences. From Eq.~(\ref{eq:Hv}), the vortex contribution decays approximately as $(h+h_m)^{-2}$. From Eq.~(\ref{eq:Hs_corrected}), the corrected image term scales as $R^3(h+\delta)^{-3}$ and increases rapidly with sphere radius. To make the nonmonotonic response arising from their competition more apparent, the calculations below use a saturation magnetization three times that of the preceding Meissner-state calculations, $M_s=5.82\times10^5~\mathrm{A/m}$.

Figure~\ref{fig:vortex}(b) first shows the vortex-free reference. Here $R=0.2~\mu\mathrm{m}$, and the vortex-free calculation gives an effective image depth $\delta\simeq76~\mathrm{nm}$. In the left panel, the order parameter is uniform except within $\xi$ of the surface and disk edge. In the right panel, the resonance frequency approaches its free-space value monotonically as $h$ increases. Agreement between the corrected image method and the numerical resonance supports the description of the Meissner term in this geometry.

Figure~\ref{fig:vortex}(c) introduces a stable single vortex with the same magnetic parameters. The left panel shows the local suppression of $|\Psi|^2$ at the core. The distance dependence of the resonance changes qualitatively in the right panel. The image term is stronger at short separations, but its $h^{-3}$ decay is faster than the $h^{-2}$ decay of the vortex term. Their relative weights therefore cross as $h$ increases, producing a nonmonotonic frequency dependence. Equation~(\ref{eq:f_vortex}), which includes both $H_v$ and $H_s'$, follows the numerical resonance, supporting this explanation in terms of competing stray-field contributions.

Finally, Fig.~\ref{fig:vortex}(d) tests this mechanism by varying the sphere radius. We compare
$R=0.1$, $0.2$, and $0.3~\mu\mathrm{m}$. The value of $\delta$ is determined independently from the vortex-free result for each radius, while $h_m=56~\mathrm{nm}$ is held fixed. Since $H_v$ is independent of $R$ whereas $H_s'\propto R^3$, the sphere radius directly controls the competition. For $R=0.1~\mu\mathrm{m}$, the vortex term dominates and the frequency decreases monotonically toward $f_0$ as the separation increases. At $R=0.2~\mu\mathrm{m}$, the terms are comparable over the range studied and the frequency first rises and then falls. At $R=0.3~\mu\mathrm{m}$, the image term dominates at short separations and lowers the frequency below its free-space value. Its faster decay then produces a monotonic increase toward $f_0$. The numerical results for all three radii follow the trends predicted by Eq.~(\ref{eq:f_vortex}).

The vortex-free and single-vortex calculations thus give a consistent physical picture. Without a vortex, the resonance is governed mainly by Meissner screening and is described quantitatively by the image method with a finite-penetration correction. A vortex adds a local stray field with a different distance dependence. Its competition with the image term produces monotonically decreasing, nonmonotonic, or monotonically increasing resonance frequencies. The formulation itself is not restricted to the high symmetry of the geometry considered. The coupling is carried entirely by the electromagnetic field, so the same approach extends to arbitrary vortex lattices, magnetic particles of arbitrary shape, multiple magnets, and continuous magnetic films with nonuniform magnetic textures.

\section{Discussion}
\label{sec:discussion}

Within the approximations adopted here, the LLG equation, the TDGL equations, and the electromagnetic equations together form a self-consistent numerical model that can test the scope of image approximations and analytical resonance formulas. At the physical level, the magnetoquasistatic approximation preserves alternating feedback but neglects electromagnetic-wave propagation and radiation, restricting the model to near-field systems much smaller than the relevant wavelength~\cite{Zhou2023Gating}. The single-domain approximation likewise excludes nonuniform spin-wave modes within the sphere. The formal validity of the conventional TDGL equations depends on proximity to the critical temperature and on the gap and relaxation mechanisms. Quantitative validity must therefore be checked away from the critical temperature or under strong nonequilibrium conditions~\cite{Schmid1966TDGL,Oripov2020TDGL}. Numerically, the computational domain and mesh control boundary-truncation errors and resolution of local structure, respectively. The coherence length, penetration depth, and vortex core must all be resolved, and the finite-disk approximation to a thick plane must be reassessed for thinner samples or spheres near an edge. At the level of the solution method, the time-domain framework retains the coupled dynamics of magnetization, order parameter, and electromagnetic field. The frequency-domain calculation instead fixes the equilibrium order parameter, keeping alternating-current feedback but omitting driven order-parameter deformation and vortex displacement. Agreement between time- and frequency-domain resonance frequencies has been verified only for the Meissner-state parameters tested. The single-vortex frequency-domain results describe a fixed superconducting background with a completely pinned vortex.

The analytical approximations have separate limits. The effective image and monopole depths are calibrated only within the range studied, and the near field of a Pearl vortex in a thin film can depart from the monopole description~\cite{Thiel2016VortexImaging}. Curved or multiple planar boundaries may require continuous image distributions or multiple images, so the analytical model must be reconstructed and tested numerically~\cite{Lin2006ImageMethod,Headley2025DipoleTrapping}. In terms of scale, the numerical model is suited to mesoscopic problems requiring simultaneous resolution of local vortices, finite penetration, and magnetization feedback. 

On the experimental side, Tian et al. observed a reduction of the resonance field below the superconducting transition in superconductor--ferromagnet multilayers and proposed an explanation based on a nonuniform Meissner field~\cite{Tian2025Meissner}. This provides a relevant experimental comparison, but the differences in geometry and in magnetic modes between thin films and spheres prevent a direct test of the present model. A concrete test could place a YIG microsphere above a Nb thin film, drive its uniform mode with a microwave cavity or coplanar waveguide, and record the resonance as a function of sphere--film separation. Preparing the vortex-free state by zero-field cooling and a pinned single-vortex state by local field-cooling, under otherwise matched geometry and bias conditions, would isolate the alternating-feedback contribution and test the predicted competition between the vortex stray field and the image term. For a direct experimental comparison, the action of the bias field on the superconductor must also be modeled, and the cooling history, vortex position, and trapped flux carefully controlled~\cite{Jeon2019Screening}.

Several extensions follow from this numerical model. Time-domain coupling could be used to study mobile vortices and multivortex states, including feedback between vortex displacement, intervortex interactions, and magnetization precession. The theory of coupled skyrmion--vortex motion provides a reference for introducing positional degrees of freedom~\cite{Menezes2019SkyrmionManipulation}. Relaxing the single-domain assumption would also allow nonuniform spin-wave responses to be studied against finite-penetration and vortex backgrounds, building on studies of spin-wave coupling to superconducting images and of hybrid spin-wave--Meissner-current modes~\cite{Golovchanskiy2018Dispersion,Borst2023Hybrid}. For multiple-interface structures such as superconductor/magnetic-insulator/superconductor stacks, theory has identified the role of electric-field reflection and Meissner-current feedback in FMR~\cite{Zhou2023Gating}. Meissner-current-mediated coupling between two ferromagnets offers another reference for extending the model to multiple magnets~\cite{Gordeeva2025Ultrastrong}. Such extensions will require the electromagnetic boundary conditions and current distributions to be reconsidered, together with the applicability of the image description and of the competing-field mechanism.

\section{Conclusions}
\label{sec:conclusion}

The numerical model developed here treats the stray-field-mediated coupling between a ferromagnet and a superconductor bidirectionally and self-consistently in both time and frequency domains. It shows that superconductivity changes magnetic resonance through both static screening and dynamic backaction. These contributions can compete, so the sign of the resonance shift cannot be inferred from the equilibrium screening field alone. Agreement between the corrected image method and numerical results shows that Meissner-state feedback admits a compact analytical description within its range of validity. In the single-vortex configuration studied, the vortex stray field and screening backaction have different distance dependences. Their competition makes the resonance sensitive to both sphere radius and height, providing an observable signature in the frequency--height relation.

These results provide a basis for controlling magnetization dynamics through superconducting screening and vortex states. Extending the numerical model to nonuniform magnetic modes and multiple vortices would allow feedback between magnetization and vortex motion to be investigated and would test how the mechanisms identified here extend to more complex superconducting boundaries. More broadly, this bidirectionally coupled micromagnetic and Ginzburg--Landau framework provides an effective tool for hybrid ferromagnet--superconductor systems in which neither subsystem can be treated as a passive background, and it provides a basis for superconducting control of magnetic excitations in hybrid magnonic structures.

\begin{acknowledgments}
This work was supported by the Fundamental and Interdisciplinary Disciplines Breakthrough Plan of the Ministry of Education of China (Grant No.~JYB2025XDXM120), the National Key Research Program of China (Grant No.~2022YFA1403300), the Innovation Program for Quantum Science and Technology (Grant No.~2024ZD0300103), and the National Natural Science Foundation of China (Grants Nos.~12574112 and 12204107). The authors thank  Chensong Hua, Ruoban Ma, Jiongjie Wang and Bakhrom Oripov for valuable discussions.
\end{acknowledgments}

\end{document}